\documentclass[twocolumn,showpacs,preprintnumbers,amsmath,amssymb]{revtex4}
\usepackage{epsfig,amssymb,euscript} 
\usepackage{amsmath}
\usepackage{mqcommands}
 \def\be{\begin{equation}}
 \def\ee{\end{equation}}
 \def\bea{\begin{eqnarray}}
 \def\eea{\end{eqnarray}}

\newcommand{\ff}{f_{\infty}}
\newcommand{\reef}[1]{(\ref{#1})}

\newcommand{\lp}{l_p}

\newcommand{\R}[1]{\mathcal R_{#1}}
\newcommand{\cb}[2]{
 \left(\begin{tabular}{c}
 #1 \\
 #2
 \end{tabular}
 \right)
 }

\begin{document}
\preprint{DAMTP-2010-37}
\title{
New massive gravity, extended.
}
\author{Miguel F. Paulos$^1$}
\email{m.f.paulos@damtp.cam.ac.uk}
\affiliation{$^1$ Department of Applied Mathematics and Theoretical Physics, Cambridge CB3 0WA, U.K.}
\date{\today}
\begin{abstract}
We consider gravity in three dimensions with an arbitrary number of curvature corrections. We show that such corrections are always functions of only three independent curvature invariants. Demanding the existence of a holographic c-theorem we show how to fix the coefficients in the action for an arbitrarily high order, recovering the new massive gravity lagrangian at quadratic order. We calculate the central charge $c$ and show that using Cardy's formula it matches the entropy of black hole solutions, which we construct. We also consider fluctuations about an AdS background, and find that it is possible to obtain two derivative equations by imposing a single constraint, thereby lifting the pathologic massive modes of new massive gravity. If we do not impose this, there is a set of ghosty massive modes propagating in the bulk. However, at $c=0$ these become massless and it is expected that these theories encode the dynamics of the spin two sector of strongly coupled logarithmic CFT's.
\end{abstract}
\maketitle

New massive gravity (NMG) \cite{Bergshoeff:2009hq, Bergshoeff:2009aq} is a theory of gravity in three dimensions with curvature squared corrections, which has been the object of much attention recently \cite{Andringa:2009yc,Bergshoeff:2009tb,Bergshoeff:2009fj}. The theory is equivalent at the linearized level to a massive spin-2 Pauli-Fierz theory. The theory is unitary in flat-space, but in AdS space the either the massless or massive graviton modes behave like ghosts \cite{Liu:2009bk}.  This theory has been recently shown to support a version of the holographic c-theorem \cite{Freedman:1999gp}, and that is has cubic and quartic curvature analogues \cite{Sinha:2010ai}. Inspired by this result, in this letter we construct extensions of new massive gravity involving an arbitrary number of curvatures, by requiring that the holographic c-theorem holds. 

\section{Higher derivative actions}

The gravitational action for NMG is given by \cite{Bergshoeff:2009hq}
\be
S=\frac{1}{\lp} \int d^{3}x\, \sqrt{-g}\left(R+\frac{2}{L^2}+4 \lambda\left( \tilde R_{ab}\tilde R^{ab}-\frac{1}{24}R^2\right)\right)\nonumber
\ee
where we have defined $\tilde R_{ab}=R_{ab}-g_{ab}R/3$, the traceless part of the Ricci tensor. We would like to generalize this action by including higher number of curvatures. Since we are in three dimensions, the curvature tensor can be written in terms of the Ricci tensor and scalar, so we only need to worry about these.
Start by defining $(\mathcal R_n)_{a}^{b}\equiv \tilde R^{b}_{\ j_1}\tilde R^{j_1}_{\ j_2}\ldots \tilde R^{j_{n-1}}_{\ a}$, and its trace $\mathcal R_n=(\mathcal R_n)_{a}^{a}$. An arbitrary curvature invariant of a given order can be conveniently written in terms of integer partitions. This can be done by establishing the dictionnary
\bea
R \to \mbf 1,\qquad \mathcal R_{n}\to \mbf n, \qquad \times\to + \nonumber
\eea
A given invariant then maps to an integer partition, and vice-versa. In this way it is possible to construct all possible invariants at a given order with ease. For instance, at order $\mbf {4}$ we have the partitions/invariants:
\bea
\mbf 1+\mbf 1+\mbf 1+\mbf 1\quad &\to&\quad  R^4 \nonumber \\
\mbf 1+\mbf 1+\mbf 2\quad &\to & \quad R^2 \mathcal R_2 \nonumber \\
\mbf 2+\mbf 2\quad &\to &\quad \mathcal (\mathcal R_2)^2 \nonumber \\
\mbf 1+\mbf 3 \quad & \to & \quad R \, \mathcal R_3 \nonumber \\
\mbf 4 \quad & \to & \quad \mathcal R_4.
\eea
Since we are in three dimensions, there are Schouten identities linking different invariants of a given order. Such identities can occur because of hidden antisymmetrizations over four indices. For instance, at quartic order we get:
\bea
0= \delta^{\ i_1 i_2 i_3 i_4}_{[j_1 j_2 j_3 j_4]} \tilde R_{i_1}^{j_2}\tilde R_{i_2}^{j_3}\tilde R_{i_3}^{j_4}\tilde R_{i_4}^{j_1}=\frac 14\mathcal R_4-\frac 18 \left(\mathcal R_2\right)^2, \label{r4id}
\eea
so that $\mathcal R_4$ is not an independent invariant.
Similarly, we can find
\bea
0&=& \delta^{\ i_1 i_2 i_3 i_4}_{[j_1 j_2 j_3 j_4]} \tilde R_{i_1}^{j_2}\tilde R_{i_2}^{j_3}\tilde R_{i_3}^{j_4}\tilde R_{i_4}^{a}R_{b}^{j_1}=\nonumber \\
&=&\frac 14 (\R5)_{b}^{a}-\frac 18 \R2(\R3)_{b}^{a}-\frac 1{12} \R3 (\R2)_{b}^{a} 
\eea
Such identities are easily found with the help of a computer algebra package \cite{Peeters:2007wn}. Noticing that $(\R{n+1})_a^b=\tilde R_{a}^{c}(\R n)_c^b$, the above show that there are no independent $\R n$ curvature invariants beyond $n=3$. Therefore, at a given order $n$ it is sufficient to consider invariants given by the integer partitions of $n$ into $1,2,3$. The number of such invariants grows like $n^2$.

With these results, we can say that the most generic higher curvature theory without derivatives of curvatures has an action given by
\be
S=\frac{1}{\lp}\int d^3 x \sqrt{-g}\left(R+\frac 2{L^2}+\sum b^i_{jk} R^{i} (\R2 )^j (\R3 )^k \right)
\ee
for some arbitrary constants $b^{i}_{jk}$. 

\section{Constraints}

We would now like to constraint the undetermined coefficients in the action such that the theory has nice properties. In particular, we would like for the theory to exhibit a holographic version of the $c$-theorem. We consider then adding a matter sector to the action above, and looking for solutions of the form
\be
ds^2=dr^2+e^{2A(r)}(-dt^2+dx^2). \label{bg}
\ee
With this background the basic invariants evaluate to
\bea
R&=&-4\left(A''+\frac 32 A'\right),\nonumber \\
\R2&=& \frac 23 (A'')^2, \quad \R3=-\frac 29 (A'')^3.\label{expr} 
\eea
Plugging these results into the action leads to a complicated expression which generically involves higher powers of $A''(r)$. We would like to impose that the equation of motion for $A(r)$ is two derivative at most. This will give constraints on the various coefficients in the action, for every order $M$. At a given order $M$, consider the lagrangian density
\be
\mathcal L^{(M)}=\hat \sum \frac{c^i_{jk}}{(-4)^i\left(\frac 23\right)^j\left(-\frac2 9\right)^k} R^{i} (\R2 )^j (\R2 )^k.
\ee
where the $\hat \ $ indicates a sum constrained by $i+2j+3k=n$. Using the expressions \reef{expr} it is easy to see that the coefficient of $(A'')^n$ in the above is given by $\sum c^{i}_{jk}$, and so we must demand that this should be zero. Similarly, one finds that the coefficient of $(A'')^{n-r}$ is given by
\be
\hat \sum \cb{i}{r} c^{i}_{jk}. \label{syst}
\ee
Finally, we fix the overall coefficient of the lagrangian by choosing the rescaling
\be
c^{n}_{00}=\lambda_n\, \mathcal N_n\equiv \lambda_n \frac{4(-1)^{n+1}}{2n-3} \left(\frac{2}{3}\right)^{n-1} 
\ee
Overall, we are imposing $n$ constraints on the various coefficients $c^{i}_{jk}$. Solving the above constraints corresponds to solving a linear system, which can be easily achieved. 
Since the number of invariants grows as $n^2$, there will be a large degeneracy at higher orders. To see why this is so, notice that already at $n=6$ the number of constraints is smaller then the number of parameters, and not all coefficients will be fixed. This corresponds to the fact that exchanging $(\R2)^3 \to (\R3)^2$ does not affect the constraints.

So far, our discussion has been completely general. It is however useful to discuss some concrete examples to clarify. Let us start first with NMG theory itself. At quadratic order there are only two independent invariants. The quadratic lagrangian density evaluates to
\bea
\mathcal L^{(2)}=c^{2}_{00} (A''+3 A'/2)^2+c^{0}_{10}(A'')^2
\eea
The constraints are $c^{2}_{00}=-8\lambda_ 2/3$ and $c^{2}_{00}+c^{0}_{10}=0$. Therefore the quadratic lagrangian becomes
\be
\mathcal L^2=4\lambda_2 \left(\R 2-\frac 1{24}R^2\right)
\ee
as required. It is also easy to see that the cubic and quartic cases lead to the same results as were found in \cite{Sinha:2010ai}. The first new example consists of the quintic theory. There are $\mathcal N_{5}=5$ independent invariants, namely
$
R^5, R^3 \R2,  R^2 \R3, R\, (\R2)^2, \R2 \R3. 
$
The constraints for this theory correspond to solving the linear system
\be
\left(\begin{tabular}{ccccc}
1 & 1& 1& 1& 1 \\
5 & 3& 2& 1& 0 \\
10 & 3& 1& 0& 0 \\
10 & 1& 0& 0& 0 \\
5 & 0& 0& 0& 0
\end{tabular}
\right)
\left(\begin{tabular}{c}
$c^{5}_{00}$ \\
$c^{3}_{10}$ \\
$c^{2}_{01}$ \\
$c^{1}_{20}$ \\
$c^{0}_{11}$
\end{tabular}
\right)
=
\left(\begin{tabular}{c}
0 \\
0 \\
0 \\
0 \\
$5 \lambda_5\, \mathcal N_5$
\end{tabular}
\right).
\ee
This can be easily solved, leading to the lagrangian
\bea
\mathcal L^{(5)}&=&-\lambda_5\left(\frac{64}{21} \R3 \R2-\frac{20}{21} R (\R2)^2 \right.\nonumber \\
&&\left.+\frac {40}{63} R^2 \R3-\frac 5{189} R^3 \R2+\frac 1{9072} R^5\right).
\eea
We shall shortly see that this theory, along with all others constrained in this fashion, satisfies a holographic $c$-theorem. To summarize, to construct the $n$-th order lagrangian one proceeds in two steps. Firstly, construct all independent invariants at that order, by using the equivalence with partitions of $n$. Once the invariants are known, one solves the system of equations \ref{syst} for the various coefficient $c^{i}_{jk}$. This is a completely algebraic, algorithmic procedure, which does not require knowledge of the metric or the evaluation of invariants. In this way it is straightforward to build arbitrarily high order lagrangians with the assistance of a computer.

\section{AdS vacua and holographic $c$-theorem}

Once all constraints are imposed, the generic lagrangian evaluated on the background $\reef{bg}$ becomes

\bea
L_A&=&-\frac{4}{\lp} \,e^{2A}\sum_n \lambda_n\, \frac{(-1)^n}{2n-3}\nonumber \times \\
&\times &\left[n\,(A')^{2n-2}A''+\frac 32 (A')^{2n}\right] \label{onshell}
\eea
The equation of motion for $A$ is then
\be
\sum_n (-1)^{n}\lambda_n [(A')^{2n}+n (A')^{2n-2}A'']=0 \label{Aeqn}
\ee
Generically the presence of a matter sector will add a source term to the right hand side. Setting first $A(r)=r \sqrt{\ff}/L$, which corresponds to an $AdS_3$ background, we obtain (with $\lambda_0=\lambda_1\equiv 1$):
\be
1-\ff+\lambda_2 \ff^2-\lambda_3 \ff^3+\ldots=0. \label{ct}
\ee
The $AdS_3$ vacuum has effective radius $\hat  L^2=L^2/\ff$, with $\ff$ satisfying the above relation.

Now consider the presence of a matter sector. The equation of motion above should now have some combination of components of the stress tensor on the right-hand side. That is, the equation above consists of some combination of components of Einsteins equations.

We are interested in the difference between the $rr$ and $tt$ equations of motion, since these satisfy
\be
(\mbox{EOM}|_r^r-\mbox{EOM}|_t^t)=T_{t}^t-T_{r}^r.
\ee
Here $T_{ab}$ is the matter stress tensor, and by the null energy condition the RHS in the above should be smaller than zero. After some work it can be shown that the above becomes:
%
\be
\sum_n (-1)^{n}\,n\,\lambda_n\, (A')^{2n-2}A''=T_{t}^t-T_{r}^r.
\ee
Define now the central charge function
\be
c(r)=\frac{L}{\lp A'} \sum \lambda_n (-1)^n \frac{n}{2n-3} (A')^{2n-2}.
\ee
Then it is easy to show using the equation of motion that $c'(r)$ is proportional to $T_{r}^{r}-T_{t}^{t}$, and therefore it is always positive if the null energy condition applies. Therefore, $c(r)$ satisfies a version of the holographic c-theorem \cite{Freedman:1999gp}, now generalized for an arbitrarily high curvature gravity theory. In the $AdS_3$ background $c$ is given by
\bea
c(r)=\hat c&=&\frac{\hat L}{\lp} \sum \lambda_n (-1)^n \frac{n}{2n-3} \ff^{n-1}\nonumber \\
&=& \frac{\hat L}{\lp}(1+2 \ff \lambda_2-\ff^2 \lambda_3+\ldots)
\eea
By evaluating our full action in a particular background \cite{Sinha:2010ai}, it is possible to find the Weyl anomaly for our theory and check that it gives exactly the result above.

\section{Black hole solutions and entropy}

In this section we consider black hole solutions. Our ansatz is given by
\be
ds^2=\frac{L^2 du^2}{4u^2 f(u)}+\frac{r_0^2}{u L^2}\left( -\frac{f(u)}{\ff} dt^2+dx^2\right).
\ee
The $AdS_3$ boundary is now located at $u=0$. The above is a solution provided that
\be
f(u)=\ff(1-u)
\ee
with $\ff$ satisfying (\ref{ct}). The background has a temperature $T=\frac{r_0}{\pi L \hat L}$

Now let us compute the entropy using Wald's formula \cite{Wald}
\be
S=\pi \left(\sqrt{g_{xx}}\frac{\delta L}{\delta R^{ut}_{\ \ ut}}\right)_{u=1}
\ee
\smallskip

The tensor $\tilde R_{ab}$ is proportional to $f''(u)$, and therefore vanishes in this background. Therefore in computing the entropy only invariants of the form $R^p$ contribute. Using that $R=-6 \ff/L^2$, the lagrangian must necessarily be of the same form as (\ref{onshell}) with $A''=0, A'=\sqrt{\ff}$. Then we get

\bea
\frac{\partial L}{\partial R}&=& -\frac 1{6\ff} \frac{\partial L_A}{\partial \ff}\bigg|_{A'=\sqrt{\ff}}=2 \sqrt{\ff}\hat c \nonumber \\
\Rightarrow S&=& \frac{\pi^2 T}{\sqrt{\ff}} \frac{\partial L}{\partial R}=2\pi^2 T \hat c
\eea
as expected for a two-dimensional CFT with central charge $\hat c$.

\section{Perturbations}
Consider once again the $AdS_3$ solution (\ref{bg}) with $A(r)=r \sqrt{\ff}$, and $\ff$ satisfying (\ref{ct}), and consider adding a perturbation of the form $g_{ab}\to g_{ab}+\kappa h_{ab}$ to the metric. Imposing traceless transverse gauge, we find after a straightforward calculation the equation of motion satisfied by the perturbation:
\be
\left[ \hat c\left ( \nabla^2+ \frac{2 \ff}{L^2}\right)+2L^2 \gamma \left( \nabla^2+ \frac{2 \ff}{L^2}\right)^2\right]h_{ab}=0
\ee
with
\bea
\gamma&=& \frac {\hat L}{2\lp}\sum \ff^{n} c^{n}_{20} \times \left(\frac 32\right)^{n+1} \nonumber \\
&=& \frac{\hat L}{\lp}\sum \lambda_n (-1)^n \ff^{n-2}\, \frac{n (n-1)}{(2n-3)}\nonumber \\
&=&\frac{\hat L}{\lp}\left(2\lambda_2-2\ff \lambda_3+\frac {12}5 \ff^2 \lambda_4+\ldots\right) \label{eqng}
\eea
By setting $\gamma$ to zero we can make the equations two derivative. Of course this is only true around the special $AdS_3$ background. The kinetic term of the graviton is then controlled by the sign of $\hat c$ in such a way that the dual CFT is unitary when there is unitarity in the bulk. Keeping $\gamma$ non-zero we can recast the equation of motion as
\bea
\left ( \nabla^2+ \frac{2 \ff}{L^2}\right)\left ( \nabla^2+ \frac{2 \ff}{L^2}+\frac{\hat c}{2L^2 \gamma}\right)h_{ab}&=&0, 
\eea
We see that there are now a new set of massive modes. Requiring that these massive modes do not break the $AdS_3$ type asymptotics leads to the constraint $\hat c/\gamma<0$. However, this does not guarantee that these modes preserve unitarity. Indeed, it is straightforward to perform the analysis of \cite{Liu:2009bk} for our generalized theory. One then finds that the massless (massive) states have energies proportional to $\hat c$ ($-\hat c$). Therefore, at least one set of states always behaves as a ghost for non-zero $\gamma$.

\section{Discussion}
In this letter we have shown how to construct an infinite class of higher derivative gravity theories. Notice that somewhat similar theories to the ones constructed here have appeared recently in the literature, but in higher dimensions \cite{Myers:2010ru,Myers:2010jv,Oliva:2010eb,Oliva:2010zd}. By imposing suitable constraints on the parameters of the lagrangian we have found that it is possible to define a holographic version of the c-theorem, and that the central charge so obtained is consistent with a calculation of entropy in a black hole background. Generically this isn't sufficient to fix all parameters. In particular, degeneracies will occur by the fact that the constraints cannot distinguish between $\R2^3$ and $\R3^2$.

The structure of these theories is intriguing. Higher curvature terms encode information on the $n$-point functions of the dual stress-tensor. In particular terms of the form $R^n$ control the size of the $AdS$ space, and $R^m \R2$ type terms control the linearized perturbation equation of motion. From a holographic perspective the latter are responsible for the non-trivial contributions to the two point function, i.e. to the appearance of massive states, as can be seen by the contributions to $\gamma$ in (\ref{eqng}) . Analogously, $R^p \R3$ should give non-trivial contributions to the three-point function. In general, $n$-point stress tensor correlators would receive non-trivial contributions from $\R n$ type terms in the action, but as we've seen these can always be recast as products of $R,\R2,\R3$. This is suggestive of the fact that in a two-dimensional CFT, the $n$-point functions of the stress-tensor are completely determined by the $2$ and $3$ point functions. 

We have found that by imposing the simple constraint $\gamma=0$ we get an infinite family of theories with two-derivative equations for linearized perturbations. In this situation there are no ghosty massive modes in the bulk, and both the gravity theory and dual conformal field theories are unitary by demanding that the central charge $\hat c$ is positive. This provides us with an infinite set of seemingly consistent toy models for quantum gravity in three-dimensions. Parameters in the lagrangian encode  $n$-point functions of the stress-tensor which are kept finite in the supergravity limit, unlike for instance what happens with $N=4$ super Yang-Mills and its Type IIB supergravity dual.

For $\gamma$ non zero, there are massive states  propagating in the bulk, but which are ghosty, as in the original NMG theory. In this case it has been argued that the at the particular point $\hat c=0$ the bulk theory can still be used to describe a logarithmic conformal field theory \cite{Skenderis:2009nt,Grumiller:2009mw,Grumiller:2009sn}. It would be interesting to compute the structure of the two point function and three point function in more detail, and in particular at the critical point  $\hat c=0$, along the lines of \cite{Grumiller:2009mw,Grumiller:2009sn}. We leave this and other questions for future work.

\acknowledgments{
It is a pleasure to acknowledge the useful comments and suggestions of Robert Myers, Aninda Sinha and Jose Edelstein. This work was supported by the Portuguese government, FCT grant SFRH/BD/23438/2005, and by DAMTP, University of Cambridge.}

\bibliography{QT3D}{}
\end{document}